\documentclass[a4paper,11pt]{article}
\usepackage{pos}
\usepackage{cleveref}
\newlength{\bibitemsep}
\newlength{\bibparskip}
\let\oldthebibliography\thebibliography
\renewcommand\thebibliography[1]{%
  \oldthebibliography{#1}%
  \setlength{\parskip}{\bibitemsep}%
  \setlength{\itemsep}{\bibparskip}%
}

\usepackage{amsmath}
\usepackage{graphicx}

\title{EUSO-SPB2 Cherenkov Telescope: Overview and First Neutrino Constraints}
\ShortTitle{EUSO-SPB2: Overview, First Neutrino Constraints}
\author[a]{Tobias Heibges}
\author[b]{Diksha Garg}
\author[c]{Claire Gu\'epin}
\author[a]{Julia Burton-Heibges}
\author[d]{John F. Krizmanic}
\author*[b]{Mary Hall Reno}
\author[d]{Tonia M. Venters}
\author[a]{Lawrence Wiencke}

\affiliation[a]{Colorado School of Mines, Department of Physics,\\
1523 Illinois Street, Golden, CO, USA}
\affiliation[b]{University of Iowa, Department of Physics and Astronomy,\\
  Iowa City, IA,  USA}
\affiliation[c]{Laboratoire Univers et Particules de Montpellier,
\\
Montpellier, France}
\affiliation[d]{Goddard Space Flight Center,\\
Greenbelt, MD, USA}
\onbehalf{for the JEM-EUSO collaboration} 

\emailAdd{theibges@mines.edu}
\emailAdd{diksha-garg@uiowa.edu}
\emailAdd{claire.guepin@lupm.in2p3.fr}
\emailAdd{julia\_burton@mines.edu}
\emailAdd{john.f.krizmanic@nasa.gov}
\emailAdd{mary-hall-reno@uiowa.edu}
\emailAdd{tonia.m.venters@nasa.gov}
\emailAdd{lwiencke@mines.edu}

\abstract{Earth-skimming tau neutrinos with energies above $\sim 10$ PeV can convert to tau leptons that decay in the atmosphere and initiate upward-going extensive air showers that generate optical Cherenkov signals. On a curtailed NASA balloon flight in May 2023, the Cherenkov telescope (CT) on the Extreme Universe Space Observatory on a Super Pressure Balloon 2 (EUSO-SPB2) was launched and had a short flight at $\sim 30$ km altitude. With some time pointing below the Earth’s limb, EUSO-SPB2 CT data allow searches for neutrino events that yield optical flashes from the forward-beamed Cherenkov light. We present an overview of the CT and provide upper limits for the diffuse astrophysical neutrino flux from flight data as a proof-of-principle demonstration. 
We also briefly describe how the methodology is extended to potential transient neutrino point sources.}

\ConferenceLogo{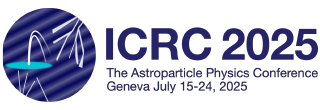}

\FullConference{%
39th International Cosmic Ray Conference (ICRC2025)\\
  15 - 24 July 2025\\
  Geneva, Switzerland}

\begin{document}

\maketitle

\section{Introduction}

\vspace{-1ex}

Ultrahigh energy neutrinos along with electromagnetic radiation and gravitational waves are messengers of some of the most energetic environments in astrophysics. The IceCube Collaboration's measurements of the diffuse astrophysical neutrino flux \cite{IceCube:2013low} opened up the field of high-energy neutrino astrophysics. Since then, the IceCube Collaboration's reports of events associated with the starburst galaxy NGC 1068 \cite{IceCube:2019cia} and the blazar TXS 0506+056 (IceCube-170922A) \cite{IceCube:2018dnn} began an era of individual transient neutrino source searches and detection. 

The diffuse astrophysical flux of neutrinos has been measured to neutrino energies of a few PeV \cite{IceCube:2025ary}. At higher energies, a flux of cosmogenic neutrinos from cosmic ray interactions with cosmic background radiation is predicted (see, e.g., ref. \cite{Aloisio:2015ega}). Efforts to detect the fluxes of both astrophysical and cosmogenic neutrinos with energies 10 PeV and above require large detection volumes. Such large volumes are achieved by instrumenting large volumes of ice or water to detect optical Cherenkov signals from charged particles produced from neutrino interactions. Detection techniques include using surface radio detectors for radio Cherenkov signals from the Askaryan effect in the medium, using surface air Cherenkov detectors, and launching  balloon-borne radio and optical Cherenkov detectors \cite{Ackermann:2022rqc}. Balloon-borne instruments can detect Earth-skimming tau neutrinos.

The Earth-skimming tau neutrino technique exploits two features of tau neutrinos. The first feature that the Earth is more transparent to tau neutrinos than other neutrino flavors because of tau neutrino regeneration. Tau neutrinos may interact to produce $\tau$-leptons which may subsequently decay, always regenerating tau neutrinos. 
The relatively short lifetime of the $\tau$-lepton and its smaller electromagnetic energy loss compared to muons make this an important effect. The second feature is that at high enough energies, $\tau$-leptons produced by $\nu_\tau$ interactions in the Earth can emerge and decay to produce upward-going extensive air showers (EASs), which are a nearly background-free signal of Earth-skimming tau neutrinos. Balloon-borne or space-based instruments capable of detecting these upward-going EASs are able to observe the large volumes needed to detect very-high energy (VHE) neutrinos from the diffuse astrophysical flux and from transient neutrino sources \cite{Reno:2019jtr,Venters:2019xwi,Venters_2021ICRC}. The Cherenkov telescope (CT) on board the Extreme Universe Space Observatory on a Super Pressure Balloon 2 (EUSO-SPB2) mission \cite{Adams:2025owi} was designed to detect optical Cherenkov radiation from EASs induced by $\tau$-lepton decays. A second telescope on EUSO-SPB2, a fluorescence telescope, was sensitive to air fluorescence signals of cosmic ray induced EAS. EUSO-SPB2 and the future POEMMA-Balloon with Radio (PBR) mission \cite{ICRC2025Eser}
(scheduled for launch in Spring 2027) are pathfinder missions that develop technical capabilities for a future satellite-based mission such as the proposed POEMMA \cite{POEMMA:2020ykm}.

\begin{figure}[tbp]
    \centering
    \includegraphics[width=0.50\textwidth]{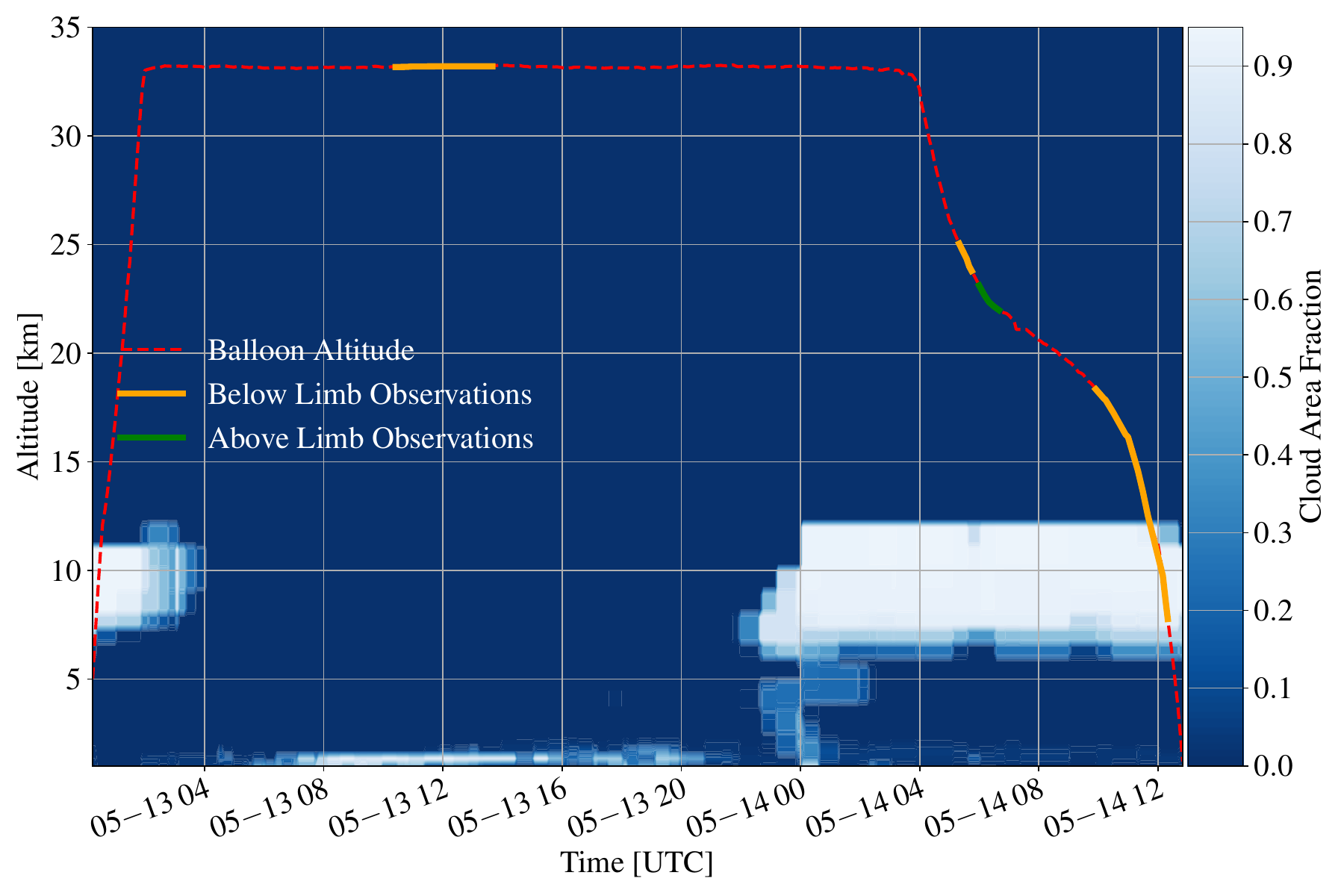}
    \caption{The red dashed line shows the balloon's altitude, while the green and orange lines indicate  observations above and below the Earth's limb, respectively. The cloud area fraction below the EUSO-SPB2 flight path is shown as a function of altitude according to the color scale on the right.}
    \label{fig:cloud-height}
\end{figure}

The EUSO-SPB2 payload, with its optical Cherenkov and fluorescence telescopes, was launched from Wanaka, New Zealand on May 13, 2023. It ascended to 33 km in altitude, and during its first night of commissioning, the Cherenkov telescope was pointed below the Earth's limb to search for upward-going EASs. During the second night, a hole in the balloon caused the payload to descend, at which point the mission was terminated. Figure \ref{fig:cloud-height} shows the balloon altitude profile for the nearly two days EUSO-SPB2 was aloft. The orange portions of the altitude curves show observation times when the Cherenkov telescope was pointing below Earth's limb. Also shown are cloud fractions as a function of altitude, discussed in  \autoref{sec:cloud-coverage}.

In these proceedings, we present an analysis of the EUSO-SPB2 data taken from observations below the Earth's limb and provide limits on the diffuse neutrino flux above $10^8$ GeV \cite{HeibgesPhDThesis2025}.  Even using a simulated 50-day flight path, limits on the VHE neutrino flux are not competitive with instruments with multiple years of data; however, the data analysis and interpretation are important for future balloon missions such as PBR and serve as a proof-of-principle demonstration for future satellite missions like POEMMA. The analysis presented here is also a precursor to understanding the instrument's sensitivity to transient neutrino sources (during target-of-opportunity, ToO, follow-up observations) where EUSO-SPB2's sensitivity is more competitive with other neutrino telescopes.

\vspace{-2ex}

\section{Cherenkov telescope, data selection and simulation criteria}
\label{sec:cloud-coverage}

\vspace{-1ex}

The Cherenkov telescope (CT) was based on a modified Schmidt design with a 1\,m diameter entrance pupil and a spherical primary mirror with a 1.64\,m radius of curvature \cite{Adams:2025owi} made up of four individual segments. The entrance pupil was an aspheric corrector lens to correct for aspheric aberrations that occur for off-axis beams due to the similar scale of the entrance pupil and the radius of curvature of the mirrors (1\,m vs. 1.64\,m).

The four CT mirror segments were aligned to divide parallel light into two beams that were separated by 1.2 cm and yield two spots of about 4-5\,mm diameter on the focal plane of the Cherenkov camera (CC). This bifocal alignment allows for the rejection of primary cosmic rays, which can trigger single pixels \cite{2023ICRCEser}. . 
Observation of a neutrino event requires two bifocal spots separated by one pixel arriving within a 10 ns time window. Simulations show that upward-going EASs have significant light deposits in the first 10 ns of detection \cite{Cummings:2020ycz}.
The bifocal split of light reduces the light intensity in a single pixel by approximately a factor of two. This has an impact on the energy threshold, increasing it by a factor $\sim 2$. 

\begin{figure}
    \centering
    \includegraphics[width=0.45\linewidth]{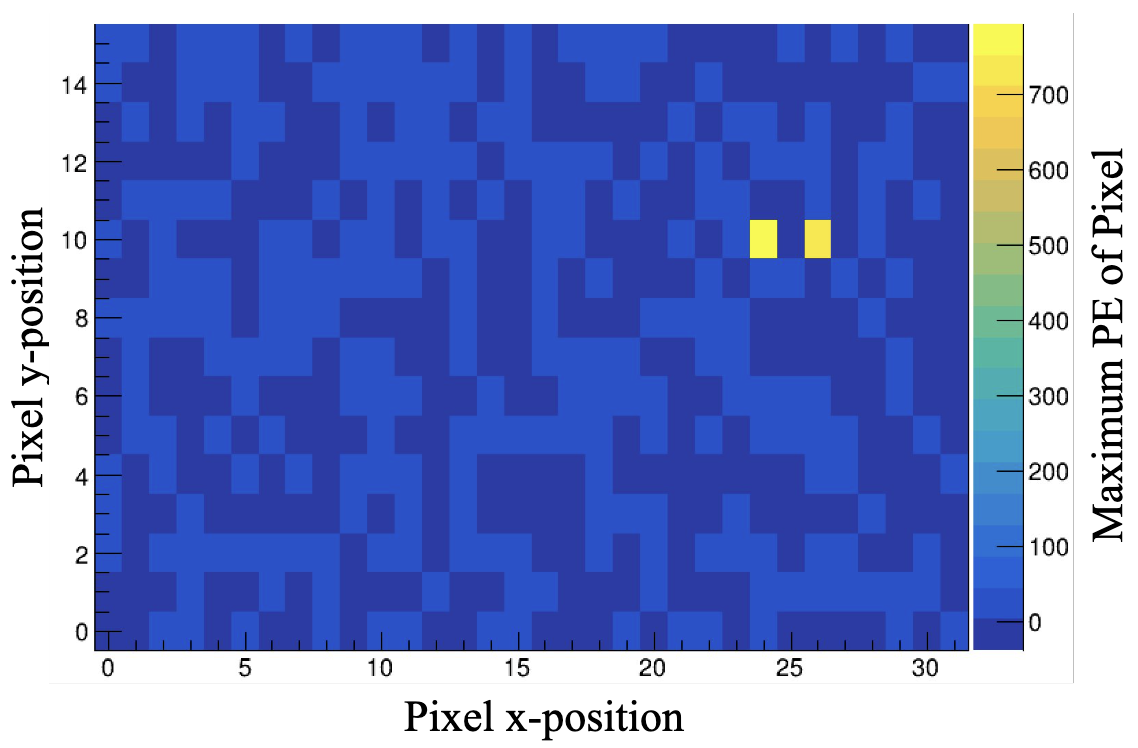}
    \caption{Simulated bifocal trigger event made using the OffLine software package \cite{JEM-EUSO:2023fyg}, neglecting camera electronics modeling, for a $5\times 10^9$ GeV EAS coming from 96$^\circ$ zenith angle (just below the Earth's limb, as seen from the balloon at 33 km), with the EAS starting after traveling 100 km distance in the atmosphere. }
    \label{fig:bifocal}
\end{figure}

The CC was comprised of 32 Hamamatsu S14521-6050AN-04 silicon photomultipliers (SiPMs), where each SiPM held $4\times 4$ pixels. Each pixel was 6\,mm$\times$6\,mm with a field of view (FoV) of $0.4^\circ\times 0.4^\circ$.
The SiPM modules were wired to an interface board with two MUSIC chips, each of which read out a set of $2\times 4$ pixels. Descriptions of the microcontroller and other electronics that control the MUSIC chips appear in ref. \cite{Bagheri:2024byu}.

The pixel configuration is shown in \autoref{fig:bifocal} with a bifocal event simulated using the OffLine software \cite{JEM-EUSO:2023fyg}, neglecting camera electronics modeling. The pixels were arranged in a $16\times 32$ pixel matrix along a curved focal plane. The CC had an FoV of $6.4^\circ$ vertical and $12.8^\circ$ horizontal and was sensitive to wavelengths between 200\,nm to 1000\,nm. The peak photon detection efficiency was $\sim 50\%$  at 450\,nm \cite{Bagheri:2024byu}.

To ensure accurate data selection and realistic simulation conditions, we
define the desired observational criteria. 
By specifying these conditions, we filtered the relevant data and determined the observation time window for our simulations.
Since the CT detects Cherenkov photons emitted by EASs, it is highly susceptible to background light. This background may arise from various sources such as sunlight, moonlight, starlight, night-sky airglow, and electronic noise. To minimize interference from celestial sources, we require astronomical night observations with either no Moon or very low moonlight. For skimming $\nu_\tau$s, only the time intervals when the CT was pointed below the Earth's limb are relevant.  
(Above-the-limb observations were performed at other times to detect high-altitude 
highly inclined 
air showers from cosmic rays \cite{ICRC2025Filippatos}.)

During the flight, unidentified artifacts in the electronics caused some of the CT pixels to overtrigger. 
There were also observation time windows during which the trigger rate reached a threshold of 500 triggers/minute, which we consider overtriggered time intervals. To maintain data quality, we excluded both the over-triggered pixels and the over-triggered time intervals. After applying these selection criteria, we obtained a total observation time of 35 minutes during Night 1. The results presented below use only Night 1 data.

During the observation time window, random noise from  night-sky background (NSB) photons was detected. Of the 512 10-ns bins per event, we use time bins 4 to 200 in the CC trace to characterize the NSB and instrument noise. These time bins are before the triggering peak, which is always localized at time bin 256.  These pre-pulse data reveal an accidental trigger rate that is not Poissonian.  Using the baseline subtracted pre-pulse data from each pixel to model the background and converting digitized SiPM amplitudes to photoelectrons (PEs), including pixel-specific voltage and temperature corrections from laboratory measurements \cite{Bagheri:2024byu}, the PE threshold is set for a background rate threshold of $10^{-7}$ Hz (one background event for the dark-sky observation time over $\sim 116$ days). The PE threshold for Night 1 is thus conservatively taken to be 66 in each of the two bifocal pixels.

An important factor to consider in the simulations is the presence and distribution of clouds since they can obscure light from EASs produced below or in the clouds. To model the cloud coverage, we used the MERRA-2 database~\cite{GMAO2015}, which provides retrospective global meteorological data from January 1980 to the near-present~\cite{MERRA2_overview}. MERRA-2 takes weather data from both satellite and ground-based observations using the National Meteorological Center’s spectral statistical interpolation method, which is a finite volume numerical technique~\cite{Parrish:1992}. The output is mapped onto a $0.5^\circ \times 0.625^\circ$ latitude-longitude grid and updated every 1, 3, or 6 hours. Depending on the observed quantity, it contains 42 or 72 vertical pressure levels and tracks parameters such as wind speed, precipitation, aerosols, and cloud cover.

\begin{figure}[!tb]
    \centering
    \includegraphics[width=0.70\textwidth]{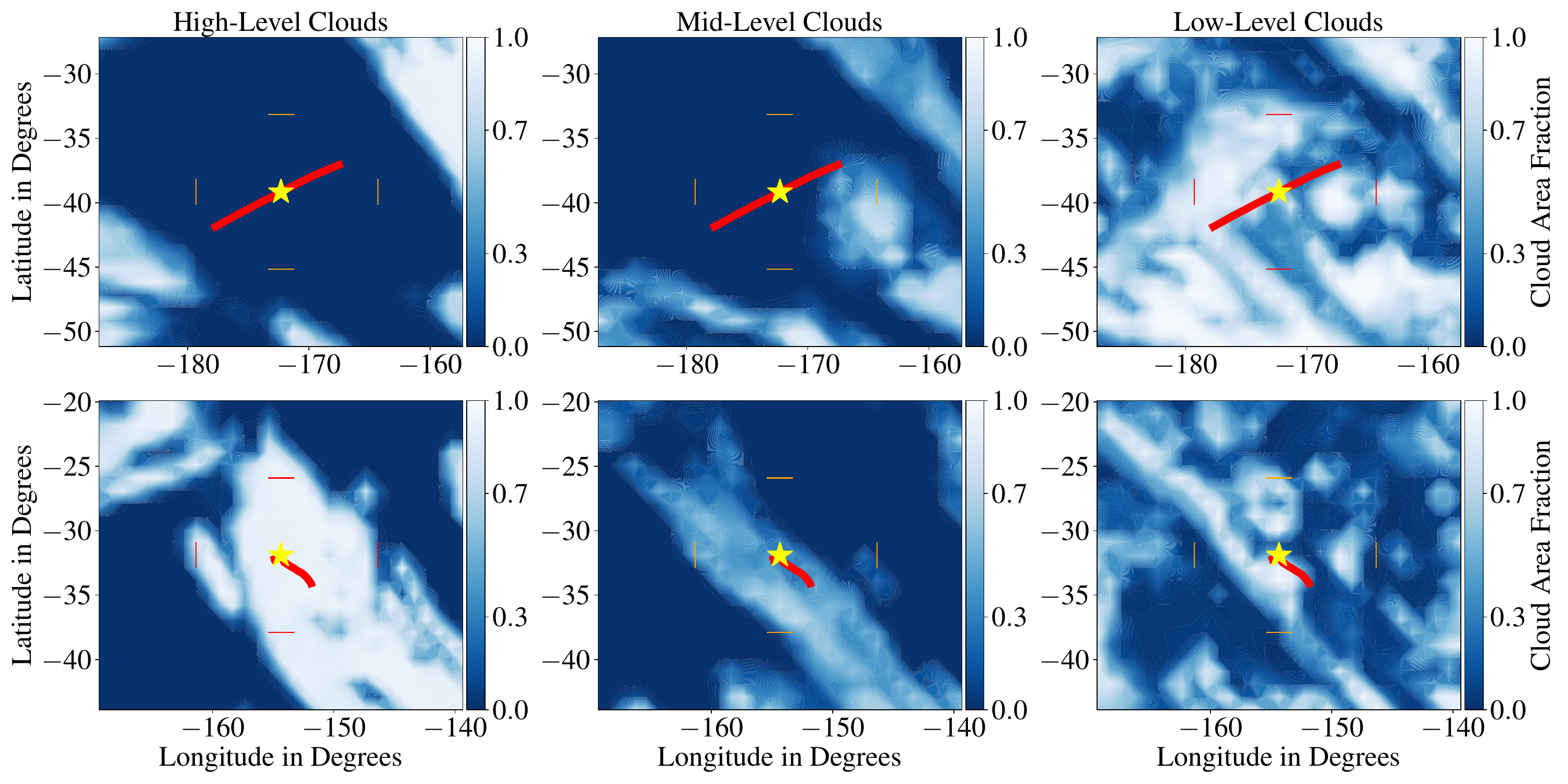}
    \caption{First, second and third columns show high-, mid-, and low-level cloud area fractions for Night 1 (top panel) and Night 2 (bottom panel). The red line indicates the balloon trajectory, and the yellow star indicates the time stamp of the cloud snapshot.  The orange lines indicate the approximate FoV of the CT.}
    \label{fig:cloud-fraction-spb2}
\end{figure}

Cloud coverage was extracted at low ($<0.7$ km), mid ($0.7-5$ km), and high altitudes ($>5$ km) for the two nights of the flight. The results are shown in~\autoref{fig:cloud-fraction-spb2}, where red lines represent the flight trajectory,the star marks the snapshot's time and location, and orange lines show the approximate FoV of the CT. 
On Night 1, a significant amount of low-level cloud coverage was present. Despite some spatial variation, we assume homogeneous cloud coverage. On Night 2, significant high-level cloud cover was observed, extending up to the Earth’s limb and again treated as uniformly distributed. Results from 
the MERRA-2 3D IAU State Meteorology Instantaneous 3-hourly (p-coord, $0.625^\circ \times 0.5^\circ$, L42), version 5.12.4, which includes cloud height data~\cite{GMAO2015}, are overlayed with flight altitude and observational periods in~\autoref{fig:cloud-height}. In our simulations of the EUSO-SPB2 acceptance of upward-going EASs from Earth-skimming $\nu_\tau$s, we require the emerging $\tau$-leptons to decay above the clouds and within the detector's FoV. For this analysis, $\tau$-leptons are required to decay above 2 km for Night 1 observations. This is a conservative estimate for neutrino detection as it neglects events where the EASs start within the clouds and continue to develop above them. 

\vspace{-1ex}

\section{Diffuse neutrino upper limits}\label{sec:diffuse-limits-analysis}

\vspace{-1ex}

With the requirements outlined in \autoref{sec:cloud-coverage}, no candidate neutrino events were observed. The minimum PE threshold ensures that no background events are observed. We use the Feldman-Cousins approach \cite{Feldman:1997qc} to determine the 90\% unified  confidence upper limit on the diffuse neutrino flux. Using the energy-dependent acceptance $\langle A \Omega\rangle(E_\nu)$, the upper limit per energy decade on the all-flavor diffuse neutrino flux can be written as (see, e.g., ref. \cite{Reno:2019jtr})
\begin{equation}
    E_\nu^2 \phi_{\nu,90\%}(E_\nu) = \frac{2.44\times N_\nu\times E_\nu}{\ln(10)\times \langle A\Omega\rangle(E_\nu)\times t_{\rm obs}}\,,
    \label{eq:aperture}
\end{equation}
where $N_\nu=3$ is the number of neutrino flavors, $t_{\rm obs}$ is the observation time of 35 minutes, and $2.44$ is Feldman-Cousins 90\% unified  confidence upper limit assuming no signal or background events.

The acceptance is evaluated by 
throwing 100,000 isotropic neutrino trajectories using the nuSpaceSim software \cite{2023ICRCKrizmanic}. Each trajectory is weighted by the $\nu_\tau\to \tau$ conversion probability, the $\tau$-lepton  energy is randomly generated from energy distributions of $\tau$-leptons emerging from the Earth, and the EAS starting altitude is randomly generated based on $\tau$-lepton decay and the trajectory geometry. Decays below the cloud altitude are excluded. 
Based on the energy of the shower (assumed to be $\sim 50$\% of the $\tau$-lepton energy), the number of photons that arrive at the CC is estimated using either the NuSpaceSim internal Cherenkov light model or by sampling from an EASCherSim software~\cite{Cummings:2020ycz, EASCherSim} shower library. OffLine simulations of the detector convert incoming photons to PEs, which are then tested against the cuts discussed in \autoref{sec:cloud-coverage}.

\begin{figure}[tbp]
    \centering
     \includegraphics[width=0.48\textwidth]{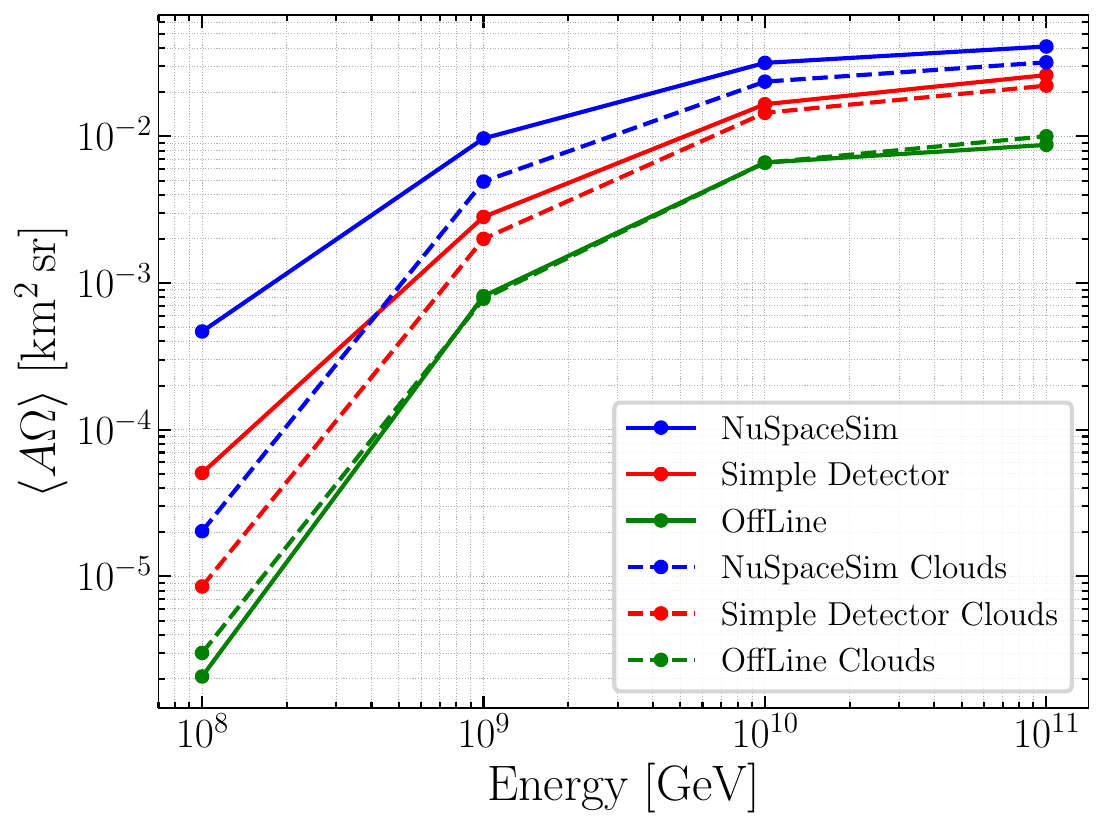}
     \includegraphics[width=0.48\textwidth]{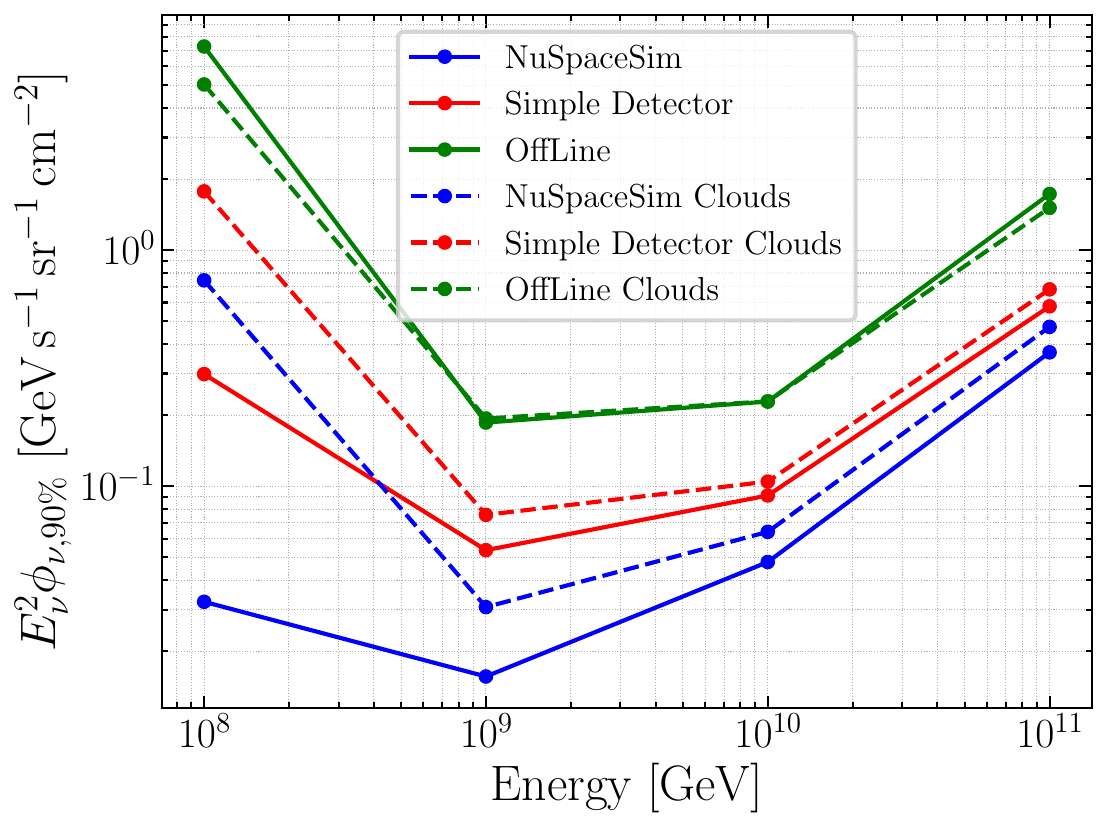}
    \caption{Left: Acceptances to diffuse $\nu_{\tau}$s as a function of energy for a 12.8$^\circ$ azimuthal FoV calculated using NuSpaceSim (blue), a simplified detector model (red), and a detailed detector model using the OffLine software (green). The solid lines represent calculations not accounting for cloud coverage, and the dashed lines represent calculations assuming clouds up to 2 km in altitude. Right: Corresponding per-decade all-flavor 90\% unified confidence upper limits on the diffuse neutrino flux as functions of energy. }
    \label{fig:diffuse-acceptance-sensitivity}
\end{figure}

We calculate the aperture using three different detector models, as shown in the left panel of \autoref{fig:diffuse-acceptance-sensitivity}. With nuSpaceSim, we model the detector as a telescope with an efficiency that is uniform across wavelength (20\% efficiency) and require the number of PEs in the camera to be greater than $2\times 66=132$ (where the factor of 2 is due to the bifocal trigger requirement). The simple detector includes a wavelength dependence in the detection efficiency and gaps in the detection surface. It also includes wavelength-dependent losses in mirror reflectivity and transmittance in the aspheric corrector plate. 
The OffLine simulation implements full ray tracing in order to provide higher-fidelity detector performance modeling. These simulations allow for asymmetric point spread functions between the two bifocal spots. It also allows for the existence of individual pixels. Due to its finite size (2-3\,mm), the point spread functions of multiple pixels can overlap, decreasing the amount of light seen in any one pixel. As demonstrated in \autoref{fig:diffuse-acceptance-sensitivity}, the detector modeling has a substantial impact on the final results at all energies, though the discrepancies are more pronounced in the $10^8-10^9$ GeV range than at higher energies.

In \autoref{fig:diffuse-acceptance-sensitivity}, we also demonstrate the impact of cloud coverage. The solid lines show the results neglecting clouds, whereas the dashed lines model the effects of clouds by requiring $\tau$-lepton decays above altitudes of 2 km. Clouds appear to have the most profound impact at the lower energies, which are more likely to decay at lower altitudes.

Using \autoref{eq:aperture}, we calculate per-decade all-flavor 90\% unified confidence upper limits on the diffuse neutrino flux, which is shown in the right panel of \autoref{fig:diffuse-acceptance-sensitivity}. Due to the limited flight time and commissioning requirements, the observation time is very short. To estimate the mission's performance during a longer flight, we used MERRA-2 wind data to model a super-pressure balloon trajectory during a hypothetical 50 day flight beginning on May 13, 2023. We used the neutrino observation scheduling (NuTargetScheduler, NuTS) software developed for this mission \cite{ICRC2025Guepin,Guepin:2025tba} and estimated that $\sim 268$~hr of dark-sky observation time would have been available for optical Cherenkov observations during such a hypothetical flight \cite{HeibgesInPrep2025}. Assuming below-the-limb observation times for $50\%$ of the 50-day dark-sky time (allowing for equal observation times pointing above and below the Earth's limb), we estimate that each of the curves in the right panel of \autoref{fig:diffuse-acceptance-sensitivity} would be multiplied by a factor of $4.4\times 10^{-3}$ in the absence of events that pass the observation requirements including the PE threshold.

More detailed discussions of the dependence of the diffuse neutrino flux sensitivity on the assumptions and approximations used here appear in refs. \cite{HeibgesPhDThesis2025,HeibgesInPrep2025}.

\vspace{-1ex}

\section{Transient neutrino sources and future prospects}

\vspace{-1ex}

For the relatively short flights of instruments on super-pressure balloons, even when aloft for 100 days, diffuse flux sensitivities like that from EUSO-SPB2 are not competitive with IceCube because the large detection volumes achievable by balloons are insufficient to make up for the exposure achievable with IceCube's more than 10 years of data, even with its smaller detection volume. The real power of a balloon-borne optical Cherenkov telescope at 33 km altitude is for searches for neutrino ToO observations. Telescopes like EUSO-SPB2 can yield competitive limits for short neutrino bursts \cite{ICRC2023Heibges}. Results from the EUSO-SPB2 data for ToO sensitivities are presented in ref. \cite{ICRC2025Guepin} for which this study of the detector characteristics is essential. The understanding of the EUSO-SPB2 instruments and the data obtained in flight will be invaluable for future balloon-borne and eventual satellite missions. PBR will be the next EUSO-type mission to be flown on a super pressure balloon and will feature a combined optical Cherenkov/fluorescence telescope augmented by radio antennas. Launch for PBR is planned for May 2027 \cite{ICRC2025Eser}.

\vskip 0.2in
\noindent{\bf Acknowledgements}
\vskip 0.2in

Work supported by NASA awards 11-APRA-0058, 16-APROBES16-0023, 17-APRA17-0066, NNX17AJ82G, NNX13AH54G, 80NSSC18K0246, 80NSSC18K0473, 80NSSC19K0626, \hfil \break 
80NSSC18K0464, 80NSSC22K1488, 80NSSC19K0627 and 80NSSC22K0426, the French space agency CNES, National Science Centre in Poland grant n. 2017/27/B/ST9/02162, and by ASI-INFN agreement n. 2021-8-HH.0 and its amendments. This research used resources of the US National Energy Research Scientific Computing Center (NERSC), the DOE Science User Facility operated under Contract No. DE-AC02-05CH11231. We acknowledge the NASA BPO and CSBF staffs for their extensive support. We also acknowledge the invaluable contributions of the administrative and technical staffs at our home institutions.


\begin{thebibliography}{10}

\bibitem{IceCube:2013low}
{\scshape IceCube} collaboration
  ~\href{https://doi.org/10.1126/science.1242856}{\emph{Science} {\bfseries
  342} (2013) 1242856} [\href{https://arxiv.org/abs/1311.5238}{{\ttfamily
  1311.5238}}].

\bibitem{IceCube:2019cia}
{\scshape IceCube} collaboration
  ~\href{https://doi.org/10.1103/PhysRevLett.124.051103}{\emph{Phys. Rev.
  Lett.} {\bfseries 124} (2020) 051103}
  [\href{https://arxiv.org/abs/1910.08488}{{\ttfamily 1910.08488}}].

\bibitem{IceCube:2018dnn}
{\scshape IceCube, Fermi-LAT, MAGIC, AGILE, ASAS-SN, HAWC, H.E.S.S., INTEGRAL,
  Kanata, Kiso, Kapteyn, Liverpool Telescope, Subaru, Swift NuSTAR, VERITAS,
  VLA/17B-403} collaboration
  ~\href{https://doi.org/10.1126/science.aat1378}{\emph{Science} {\bfseries
  361} (2018) eaat1378} [\href{https://arxiv.org/abs/1807.08816}{{\ttfamily
  1807.08816}}].

\bibitem{IceCube:2025ary}
  {\scshape IceCube} collaboration ~
  \href{https://doi.org/10.1103/2hnq-1fsx}{{\emph{Phys. Rev. D} {\bfseries 112} (2025) 012022}}
  \href{https://arxiv.org/abs/2502.19776}{{\ttfamily [2502.19776]}}.

\bibitem{Aloisio:2015ega}
R.~Aloisio et~al.
  ~\href{https://doi.org/10.1088/1475-7516/2015/10/006}{\emph{JCAP} {\bfseries
  10} (2015) 006} [\href{https://arxiv.org/abs/1505.04020}{{\ttfamily
  1505.04020}}].

\bibitem{Ackermann:2022rqc}
M.~Ackermann et~al.
  ~\href{https://doi.org/10.1016/j.jheap.2022.08.001}{\emph{JHEAp} {\bfseries
  36} (2022) 55} [\href{https://arxiv.org/abs/2203.08096}{{\ttfamily
  2203.08096}}].

\bibitem{Reno:2019jtr}
M.H.~Reno, J.F.~Krizmanic and T.M.~Venters
  ~\href{https://doi.org/10.1103/PhysRevD.100.063010}{\emph{Phys. Rev. D}
  {\bfseries 100} (2019) 063010}.

\bibitem{Venters:2019xwi}
T.M.~Venters et~al.
  ~\href{https://doi.org/10.1103/PhysRevD.102.123013}{\emph{Phys. Rev. D}
  {\bfseries 102} (2020) 123013}
  [\href{https://arxiv.org/abs/1906.07209}{{\ttfamily 1906.07209}}].

\bibitem{Venters_2021ICRC}
T.M.~Venters, M.H.~Reno and J.F.~Krizmanic ~{\emph{PoS} {\bfseries ICRC2021}
  (2021) 977}.

\bibitem{Adams:2025owi}
J.H.~Adams et~al. ~ \href{https://arxiv.org/abs/2505.20762}{{\ttfamily
  [2505.20762]}}.

\bibitem{ICRC2025Eser}
J.~Eser, A.V.~Olinto, G.~Osteria et~al.
  ~\href{https://doi.org/tbd/tbd}{\emph{PoS} {\bfseries ICRC2025} (2025) 249}.

\bibitem{POEMMA:2020ykm}
{\scshape POEMMA} collaboration
  ~\href{https://doi.org/10.1088/1475-7516/2021/06/007}{\emph{JCAP} {\bfseries
  06} (2021) 007} [\href{https://arxiv.org/abs/2012.07945}{{\ttfamily
  2012.07945}}].

\bibitem{HeibgesPhDThesis2025}
T.~Heibges, \emph{{First Results from the Cherenkov Telescope Flown on
  EUSO-SPB2}}, Ph.D. thesis, {{Colorado School of Mines}}, 2025.

\bibitem{2023ICRCEser}
J.~{Eser} et~al. ~\href{https://doi.org/10.22323/1.444.0397}{\emph{PoS}
  {\bfseries ICRC2023} (2023) 0397}.

\bibitem{Cummings:2020ycz}
A.L.~Cummings, R.~Aloisio and J.F.~Krizmanic
  ~\href{https://doi.org/10.1103/PhysRevD.103.043017}{\emph{Phys. Rev. D}
  {\bfseries 103} (2021) 043017}.

\bibitem{JEM-EUSO:2023fyg}
{\scshape JEM-EUSO} collaboration
  ~\href{https://doi.org/10.1088/1748-0221/19/01/P01007}{\emph{JINST}
  {\bfseries 19} (2024) P01007}
  [\href{https://arxiv.org/abs/2309.02577}{{\ttfamily 2309.02577}}].

\bibitem{Bagheri:2024byu}
M.~Bagheri et~al.
  ~\href{https://doi.org/10.1016/j.nima.2024.169999}{\emph{Nucl. Instrum. Meth.
  A} {\bfseries 1070} (2025) 169999}
  [\href{https://arxiv.org/abs/2406.08274}{{\ttfamily 2406.08274}}].

\bibitem{ICRC2025Filippatos}
G.~Filippatos et~al. ~\href{https://doi.org/tbd/tbd}{\emph{PoS} {\bfseries
    ICRC2025} (2025) 255}.

\bibitem{GMAO2015}
{Global Modeling and Assimilation Office (GMAO)}. Greenbelt, MD, USA: Goddard
  Space Flight Center Distributed Active Archive Center (GSFC DAAC), Accessed:
  December 2023 at doi: 10.5067/QBZ6MG944HW0", 2015.

\bibitem{MERRA2_overview}
R.~Gelaro et~al.
  ~\href{https://doi.org/https://doi.org/10.1175/JCLI-D-16-0758.1}{\emph{Journal
  of Climate} {\bfseries 30} (2017) 5419 }.

\bibitem{Parrish:1992}
D.F.~Parrish and J.C.~Derber ~{\emph{Mon. Wea. Rev.} {\bfseries 120} (1992)
  1747}.

\bibitem{Feldman:1997qc}
G.J.~Feldman and R.D.~Cousins
  ~\href{https://doi.org/10.1103/PhysRevD.57.3873}{\emph{Phys. Rev. D}
  {\bfseries 57} (1998) 3873}
  [\href{https://arxiv.org/abs/physics/9711021}{{\ttfamily physics/9711021}}].

\bibitem{2023ICRCKrizmanic}
J.F.~{Krizmanic} et~al. ~\href{https://doi.org/10.22323/1.444.1110}{\emph{PoS}
  {\bfseries ICRC2023} (2023) 1110}.

\bibitem{EASCherSim}
A.~Cummings, R.~Aloisio, J.~Eser and J.~Krizmanic
  ~\href{https://doi.org/10.1103/PhysRevD.104.063029}{\emph{Phys. Rev. D}
  {\bfseries 104} (2021) 063029}.

\bibitem{ICRC2025Guepin}
T.~Heibges, C.~Gu\'epin, T.M.~Venters et~al.
  ~\href{https://doi.org/tbd/tbd}{\emph{PoS} {\bfseries ICRC2025} (2025) 1051}.

\bibitem{Guepin:2025tba}
  T.~Heibges, C.~Gu\'epin et~al.
   [\href{https://arxiv.org/abs/2509.13844}{{\ttfamily 2509.13844]}}.

\bibitem{HeibgesInPrep2025}
{\scshape JEM-EUSO} collaboration ~{\emph{in preparation} (2025) }.

\bibitem{ICRC2023Heibges}
{\scshape JEM-EUSO} collaboration
  ~\href{https://doi.org/10.22323/1.444.1134}{\emph{PoS} {\bfseries ICRC2023}
  (2023) 1134} [\href{https://arxiv.org/abs/2310.12310}{{\ttfamily
  2310.12310}}].

\end{thebibliography}

\providecommand{\href}[2]{#2}\begingroup\raggedright\endgroup


    \newpage
{\Large\bf Full Authors list: The JEM-EUSO Collaboration}

\begin{sloppypar}
{\small \noindent
M.~Abdullahi$^{ep,er}$              
M.~Abrate$^{ek,el}$,                
J.H.~Adams Jr.$^{ld}$,              
D.~Allard$^{cb}$,                   
P.~Alldredge$^{ld}$,                
R.~Aloisio$^{ep,er}$,               
R.~Ammendola$^{ei}$,                
A.~Anastasio$^{ef}$,                
L.~Anchordoqui$^{le}$,              
V.~Andreoli$^{ek,el}$
% ...
}
\end{sloppypar}
\vspace*{.3cm}

{ \footnotesize
\noindent
%
% ...
$^{cb}$ Universit\'e de Paris, AstroParticule et Cosmologie (CNRS), Paris, France\\
% ...
$^{ef}$ Istituto Nazionale di Fisica Nucleare (INFN), Sezione di Napoli, Naples, Italy\\
% ...
$^{ei}$ Istituto Nazionale di Fisica Nucleare (INFN), Sezione di Roma Tor Vergata, Rome, Italy\\
% ...
$^{ek}$ Istituto Nazionale di Fisica Nucleare (INFN), Sezione di Torino, Turin, Italy\\
$^{el}$ Universit\`a di Torino, Dipartimento di Fisica, Turin, Italy\\
% ...
$^{ep}$ Gran Sasso Science Institute (GSSI), L’Aquila, Italy\\
$^{er}$ Istituto Nazionale di Fisica Nucleare (INFN), Laboratori Nazionali del Gran Sasso, Assergi, Italy\\
% ...
$^{ld}$ University of Alabama in Huntsville, Huntsville, AL, USA\\
$^{le}$ City University of New York (CUNY), Lehman College, Bronx, NY, USA\\
% ...
}

\end{document}